\documentclass[prd, aps, twocolumn, showpacs,superscriptaddress,nofootinbib]{revtex4-1}

\usepackage{amsbsy,amsmath,amssymb,amsthm,wasysym,amsfonts}
\usepackage{graphicx}
\usepackage{psfrag}
\usepackage{epstopdf}
\usepackage{color}
\usepackage{mathrsfs}
\usepackage{comment}
\usepackage[normalem]{ulem}
\DeclareSymbolFontAlphabet{\mathrsfs}{rsfs}
\DeclareMathAlphabet\mathbfcal{OMS}{cmsy}{b}{n}

\newcommand{\be}{\begin{equation}}  
\newcommand{\ee}{\end{equation}}
\newcommand{\bea}{\begin{eqnarray}}           
\newcommand{\eea}{\end{eqnarray}} 
\newcommand{\beqn}{\begin{eqnarray*}}
\newcommand{\eeqn}{\end{eqnarray*}}
\newcommand{\ba}{\begin{align}}
\newcommand{\ea}{\end{align}}

\def\de{\partial}
\def\lm{{\ell m}}   

\def\p4{{\psi_4}} 

\usepackage{float}
\definecolor{cyan}{rgb}{0,0.9,0.9}
\definecolor{orange}{rgb}{0.9,0.5,0}
\definecolor{magenta}{rgb}{1,0,1}
\definecolor{purple}{rgb}{0.8,0.4,0.8}

\definecolor{dodgerblue}{rgb}{0.12, 0.56, 1.0}
\definecolor{alizarincrimson}{rgb}{0.82, 0.1, 0.26}

\begin{document}


\title{Parametrized-4.5PN TaylorF2 approximant(s) and tail effects to quartic nonlinear order from the effective one body formalism.}

\author{Francesco \surname{Messina}}
\affiliation{Dipartimento di Fisica Universit\`a di Torino, Via P.~Giuria 1, 10125 Torino, Italy}
\author{Alessandro \surname{Nagar}}
\affiliation{Centro Fermi - Museo Storico della Fisica e Centro Studi e Ricerche ``Enrico Fermi'', 00184, Roma, Italy}
  \affiliation{INFN Sezione di Torino, Via P.~Giuria 1, 10125 Torino, Italy}
  \affiliation{Institut des Hautes Etudes Scientifiques, 91440
  Bures-sur-Yvette, France}
\begin{abstract}
  By post-Newtonian (PN) expanding the well-known, factorized and resummed,
  effective-one-body energy flux for circularized binaries we show that:
  (i) because of the presence of the resummed tail factor, the 4.5PN-accurate
  tails-of-tails-of-tails contribution  to the energy flux recently computed
  by Marchand et al. [Class. Q. Grav. 33 (2016) 244003] is  actually contained
  in the resummed expression; this is also the case of the the next-to-leading-order
  tail-induced spin-orbit term of Marsat et al. [Class. Q. Grav. 31 (2014) 025023];
  (ii) in performing this expansion, we also obtain, for the first time, the
  explicit 3.5PN leading-order tail-induced spin-spin flux term;
  (iii) pushing the PN expansion of the (nonspinning) EOB flux up
  to 5.5PN order, we compute 4PN, 5PN and 5.5PN contributions to
  the energy flux, though in a form that explicitly depends on,
  currently unknown, 4PN and 5PN non-test-mass corrections to the
  factorized waveform amplitudes. Within this (parametrized) 4.5PN
  accuracy, we calculate the Taylor~F2 approximant.
  Focusing for simplicity on the nonspinning case and using
  the numerical-relativity calibrated IMRPhenomD waveform
  model as benchmark, we demonstrate that it is possible to
  reproduce the derivative of the IMRPhenomD phase (say up to
  the frequency of the Schwarzschild last-stable-orbit)
  by flexing only a 4PN ``effective'' waveform amplitude
  parameter. A preliminary analysis also illustrates that
  similar results can be obtained for the spin-aligned case
  provided only the leading-order spin-orbit and spin-spin
  terms are kept. Our findings suggest that this kind of,
  EOB-derived, (parametrized), higher-order, PN approximants
  may serve as promising tools to construct Inspiral-Merger-Ringdown
  phenomenological models or even to replace the standardly
  used 3.5PN-accurate TaylorF2 approximant in searches of
  small-mass binaries. 
\end{abstract}

\date{\today}


\maketitle

\section{Introduction}

Effective-one-body (EOB)~\cite{Buonanno:1998gg,Buonanno:2000ef,Damour:2001tu,Buonanno:2005xu}
waveforms informed by numerical relativity~(NR)
simulations~\cite{Damour:2002qh,Damour:2007xr,Damour:2007vq}
have played a central role in the detection, subsequent
parameter-estimation~\cite{TheLIGOScientific:2016wfe}
analyses and GR tests~\cite{TheLIGOScientific:2016src}
of the gravitational-wave (GW) observations GW150914~\cite{Abbott:2016blz} and
GW151226~\cite{Abbott:2016nmj,TheLIGOScientific:2016pea} announced in 2015.
EOB waveforms have also been employed to build frequency-domain, phenomenological models
for the inspiral, merger and ringdown stages of the binary black hole (BBH)
coalescence~\cite{Khan:2015jqa}.
Those models were also used to infer the properties~\cite{TheLIGOScientific:2016wfe}
and carry out tests~\cite{TheLIGOScientific:2016src} of GR with GW150914 and GW151226.
Despite the success of the EOBNR enterprise, one has to remember that
EOB waveforms do not currently cover the full range of total
masses to which the LIGO/Virgo detectors might be sensitive.
In fact, for small-mass binaries, with $M<4M_\odot$ where $M=m_1+m_2$
is the total mass of the system, searches are done using
post-Newtonian (PN) approximants, notably the Taylor~F2 frequency
domain one constructed using the stationary phase approximation
(SPA, see e.g. Ref.~\cite{Buonanno:2009zt} and references therein).
This approximant is computed at 3.5PN total accuracy, involving
also spin-orbit and spin-spin terms. In addition, the 3.5PN-accurate
Taylor~F2 approximant is also used within the construction of
the inspiral part of Inspiral-Merger-Ringdown phenomenological
waveforms models, like IMRPhenomD~\cite{Khan:2015jqa}: its phasing
is modified by the addition of effective parameters that are
fitted to EOB-NR hybrid waveforms. By contrast, EOB models
incorporate 4PN-accurate orbital
information~\cite{Bini:2013zaa,Damour:2016abl,Damour:2015isa,Damour:2014jta,Bernard:2015njp,Bernard:2016wrg},
some test-particle information up to 5PN in the waveform~\cite{Damour:2009kr},
the spin-spin interaction is incorporated in a special resummed
form~\cite{Damour:2014sva,Balmelli:2015zsa} as well as
tail effects~\cite{Damour:2007xr,Damour:2008gu}.
In practice, since the EOB model actually employs analytical
information that is at {\it higher PN-accuracy} than the 3.5PN 
(and in fact for some parts, like the tail factor, it is using
infinite PN accuracy), and since the PN-expansion {\it is contained}
within the EOB model, it is actually possible to derive PN approximants
{\it beyond} the 3.5PN order, though with some parameters
that account for those specific terms (that depend on the
symmetric mass ratio only $\nu\equiv\mu/M= m_1m_2/(m_1+m_2)^2$) that 
are not fully known yet in PN theory). These additional
parameters, that are inspired by EOB knowledge, could be
then be {\it informed} by comparison with long-inspiral EOB 
or NR waveforms. A flavor of this idea was given already in 
Ref.~\cite{Damour:2012yf}, that computed the tidal part of the 
Fourier domain phase at 2.5PN accuracy, including
tail terms, but with the explicit dependence on a (still) unknown 2PN 
tidal waveform term. Once these, parametrized (EOB-NR informed), 
new approximants are obtained,  one may then ask whether 
they are any better than the plain 3.5PN one during the inspiral 
and, in particular, whether they might be reliable 
(e.g., against NR calibrated NR waveforms) also for 
searches with larger values of $M$. If it were the case, 
one could exploit improved, computationally  inexpensive, 
PN approximants beyond the mass range considered up to now.

The main purpose of this paper is to derive, from EOB first
principles, families of parametrized, higher-PN, approximant,
in order to give a general framework in which this problem can
be properly addressed. In practice, we present here two results. 
On the one hand, starting from the well-known, factorized and resummed, 
EOB energy flux and PN-expanding it up to 5.5PN order (for the nonspinning case),
we recover several high-order tail terms (notably the 4.5PN one)
in the energy flux of circularized binaries recently computed
using the Multipolar-Post-Minkowskian (MPM) formalism~\cite{Marchand:2016vox,Marsat:2013caa}
In doing so, we also obtain, for the first time at best of
our knowledge, the leading-order spin-spin tail-induced
term in the energy flux, finding it perfectly consistent
with the test-mass results. Then, focusing on the nonspinning
case for simplicity, we deliver a new 4.5PN-accurate TaylorF2
approximant in parametrized form, where the (two) parameters
account for yet uncalculated 4PN-accurate waveform information,
and we briefly investigate its flexibility and properties by 
comparison with the IMRPhenomD phenomenological model of Ref.~\cite{Khan:2015jqa}.

The paper is organized as follows: In Sec.~\ref{sec:55PN} we obtain
this parametrized 5.5PN accurate (nonspinning) flux and notably the
4.5PN tail term of Ref.~\cite{Marchand:2016vox}. We note, in passing,
that this term is part of the EOB resummed formalism from 2008~\cite{Damour:2008gu,Damour:2009kr},
though it was never written down explicitly. Section~\ref{sec:FtailNLO}
focuses on tail-induced spin terms: the NLO spin-orbit-induced tail term
of Ref.~\cite{Marsat:2013caa} is obtained explicitly and, in addition,
we similarly write down the LO spin-spin tail-induced term;
Section~\ref{sec:F2nonspinning} illustrates the parametrized 4.5PN
TaylorF2 approximant and gives an extensive comparison with the
IMRPhenomD model. A very preliminary investigation of the spinning
case is also discussed. Finally, we summarize our findings
in Sec.~\ref{sec:conclusions}. If not otherwise specified, we use
units where $c=G=1$.

\section{Parametrized 5.5PN,  nonspinning flux
         from the effective-one-body framework}
\label{sec:55PN}
The aim of this Section is to compute the energy flux for nonspinning binaries 
at 5.5PN order by Taylor expanding the corresponding, factorized and resummed, 
EOB energy flux~\cite{Damour:2008gu,Damour:2009kr}. 
Evidently, since the $\nu$-dependent information is incomplete 
both at 4PN and at 5PN order, the final result will be obtained in {\it parametrized} form, 
with specific parameters, that have clear mining within the EOB framework, 
accounting for this yet uncalculated, $\nu$-dependent, information.
In performing this calculation, we will also obtain as a bonus, and independently,
the 4.5PN tail term recently computed  by Marchand 
et al.~\cite{Marchand:2016vox} using the MPM formalism.

Going into detail, let us first recall the structure of the EOB energy flux.
It is written as a sum of multipoles as
\be
\label{eq:flux}
{\cal F}=\sum_{\ell=2}^{\infty}\sum_{\ell =-m}^{m} F^{\rm Newt}_{\lm}\hat{F}_{\lm}
\ee
where  $F^{\rm Newt}_{\ell m}$ is the usual Newtonian (leading-order) 
contribution and $\hat{F}_{\ell m}$ is the post-Newtonian correction.
Within the EOB framework, this latter  is written in a
special factorized and resummed  form as~\cite{Damour:2008gu}
\be
\label{eq:hF}
\hat{F}_{\ell m} = \left(S^{(\epsilon)}_{\rm eff}\right)^{2}|T_{\ell m}|^{2}(\rho_{\ell m})^{2\ell},
\ee
where $S^{(\epsilon)}_{\rm eff}$ is the effective source, that is the effective
EOB energy $\hat{E}_{\rm eff}(x)\equiv E_{\rm eff}/\mu$ when $\epsilon=0$  ($\ell+m$=even)
or the Newton-normalized orbital angular  momentum when $\epsilon=1$ ($\ell+m$=odd). 
The (complex) tail factor $T_{\ell m}$ resums an infinite number of leading
order logarithms. See Refs.~\cite{Damour:2008gu,Faye:2014fra}.
Starting from Eqs.~\eqref{eq:flux}-\eqref{eq:hF}, by means of a straightforward
calculation one can show that the 4.5PN term of~\cite{Marchand:2016vox}
is exactly contained in these expressions. 
The square modulus of the tail factor reads
\be
\label{eq:ModT}
|T_{\ell m}(x)|^{2}=\dfrac{4\pi E m x^{3/2}\prod_{s=1}^{\ell}\left(s^{2}+(2 E m x^{3/2}\right)^{2}}{(\ell!)^{2}
  \left(1-e^{-4\pi m E}\right)},
\ee
where $x=(G M\Omega/c^3)^{2/3}$ and $E$ is the energy of the system.
In a dynamical EOB evolution, $E$ is given by the EOB Hamiltonian
evaluated  along the dynamics (and in fact the orbital frequency parameter $x$ 
is actually replaced by the orbital velocity $v_{\varphi}^{2}$). 
Here, $E$  is given by its PN expansion along circular orbit.
Dropping for the moment the spin-dependence, that will be discussed in the next section,
to obtain the PN-expanded energy flux at 5PN we need to retain $E$ and $S^{(\epsilon)}_{\rm eff}$
along circular orbits up to 4PN order. The PN expansions
of $E(x)$ along circular orbits reads
(see Refs.~\cite{Damour:2014jta,Damour:2015isa,Damour:2016abl,Bernard:2016wrg})
\begin{align}
&E(x)=1-\dfrac{\nu}{2}x+\left(\frac{\nu ^2}{24}+\frac{3 \nu}{8}\right) x^2 +\left(\frac{\nu ^3}{48}-\frac{19 \nu ^2}{16}+\frac{27 \nu }{16}\right) x^3\nonumber\\
&+\left[\frac{35 \nu ^4}{10368}+\frac{155 \nu ^3}{192}+\left(\frac{205 \pi ^2}{192}-\frac{34445}{1152}\right) \nu^2+\frac{675 \nu }{128}\right] x^4\nonumber\\
%
&+\Bigg[-\frac{77 \nu ^5}{62208}-\frac{301 \nu ^4}{3456}+\left(\frac{498449}{6912}-\frac{3157 \pi ^2}{1152}\right) \nu ^3\\
&               +\left(\frac{123671}{11520}-\frac{9037 \pi ^2}{3072}
               -\frac{448}{15}{\rm eulerlog}_2(x)\right)\nu^2-\frac{3969}{256}\nu\Bigg]x^5\nonumber,                                       
\end{align}
where ${\rm eulerlog}_m(x)\equiv \gamma_{\rm E}+\log(2)+\dfrac{1}{2}\log(x)+\log(m)$ and $\gamma_E$ is the Euler constant.

The effective EOB energy along circular orbits $\hat{E}_{\rm eff}(x)$ is obtained
from the EOB mapping between the real and effective Hamiltonians~\cite{Buonanno:1998gg}
$E(x)=M\sqrt{1+2\nu(\hat{E}_{\rm eff}-1)}$ and reads
\begin{align}
   &\hat{E}_{\rm eff} (x)= 1-\frac{x}{2}+\left(\frac{\nu }{6}+\frac{3}{8}\right) x^2+\left(\frac{27}{16}-\frac{11 \nu }{8}\right) x^3\nonumber\\
   &+\left(-\frac{\nu ^3}{162}+\frac{17 \nu ^2}{12}+\left(\frac{205 \pi ^2}{192}-\frac{4417}{144}\right) \nu+\frac{675}{128}\right) x^4\nonumber\\
   &+\Bigg[-\frac{\nu ^4}{486}-\frac{115 \nu ^3}{216}+\left(\frac{74899}{864}-\frac{943 \pi ^2}{288}\right) \nu ^2\\
   &+\bigg(\frac{50293}{5760}-\frac{9037 \pi ^2}{3072}-\frac{448}{15}{\rm eulerlog}_2(x)\bigg)\nu
    -\frac{3969}{256}\Bigg]x^{5}\nonumber.
\end{align}
Finally, we also need the Newton-normalized orbital angular momentum $\hat{j}=\sqrt{x}j(x)$ 
along circular orbits as function of $x$. It reads (see Appendix for a reminder
of its derivation as well as Eq.~(5.4b) of Ref.~\cite{Damour:2014jta}) 
\begin{align}
  \label{eq:jorb}
&\hat{j}(x)= 1 +\left(\frac{\nu}{6}+\frac{3}{2}\right) x 
+\left(\frac{\nu ^2}{24}-\frac{19 \nu }{8}+\frac{27}{8}\right) x^2\nonumber\\
&+\left(\frac{7 \nu ^3}{1296}+\frac{31 \nu
   ^2}{24}+\left(\frac{41 \pi ^2}{24}-\frac{6889}{144}\right) \nu +\frac{135}{16}\right)x^3\nonumber\\
 &+ x^4 \bigg[-\frac{55 \nu ^4}{31104}-\frac{215 \nu
   ^3}{1728}+\left(\frac{356035}{3456}-\frac{2255 \pi ^2}{576}\right) \nu^2\nonumber\\
   &+\left(\frac{98869}{5760}-\frac{6455 \pi ^2}{1536}-\dfrac{128}{3}{\rm eulerlog}_2(x)\right)\nu+\frac{2835}{128}\bigg].
\end{align}
Then, we list the functions $\rho_{\ell m}(x)$ at the order needed for obtaining the 5.5PN flux. Up to 3PN accuracy, the $\ell=2$ and  $\ell=3$, $m=1$ and $\ell=m=3$
are the only contributions fully known, these latter recently obtained in Ref.~\cite{Faye:2014fra}. For higher PN terms, the $\nu$-dependent part is currently unknown,
but the test-particle ($\nu=0$) contributions are known. We write them explicitly, while the unknown $\nu$-dependent terms are indicated with various coefficients.
\begin{widetext}
\begin{align}
\rho_{22}(x;\nu)&=\rho_{22}^{\rm 3PN}(x;\nu)+\left(-\dfrac{387216563023}{160190110080} +\dfrac{ 9202}{2205} + {\rm eulerlog}_{2}(x)+\nu c_{22}^{\rm 4PN}\right)x^{4}\nonumber\\
                          &+ \left(-\dfrac{16094530514677}{533967033600} + \dfrac{439877 }{55566}{\rm eulerlog}_{2}(x)+\nu c_{22}^{\rm 5PN}(\nu)\right)x^{5}\\
\rho_{21}(x;\nu)&=\rho_{21}^{\rm 2PN}(x;\nu)+\biggl(\dfrac{7613184941}{2607897600} - \dfrac{107}{105}{\rm eulerlog}_{1}(x)+\nu c_{21}^{\rm 3PN}(\nu)\biggr)x^{3}\\\nonumber
&+\biggl(-\frac{1168617463883}{911303737344}+\frac{6313}{5880}{\rm eulerlog_1}(x)+\nu c_{21}^{\rm 4PN}(\nu)\biggr)x^4,\\
\rho_{33}(x;\nu)&=\rho_{33}^{\rm 3PN}(x;\nu) + \left(-\dfrac{57566572157}{8562153600} + \dfrac{13}{3}{\rm eulerlog}_{3}(x)+\nu c_{33}^{\rm 4PN}(\nu)\right)x^{4},\\
\rho_{32}(x;\nu)&=\rho_{32}^{\rm 2PN}(x;\nu) + \left(\dfrac{5849948554}{940355325} - \dfrac{104}{63}{\rm eulerlog}_{2}(x)+\nu c_{32}^{\rm 3PN}(\nu)\right)x^{3},\\
\rho_{31}(x;\nu)&=\rho_{31}^{\rm 3PN}(x;\nu)+\left(\dfrac{2606097992581}{4854741091200}+\dfrac{169}{567}{\rm eulerlog}_{1}(x)+\nu c^{\rm 4PN}_{31}(\nu)\right)x^{4},\\
\rho_{44}(x;\nu)&=\rho_{44}^{\rm 2PN}(x;\nu)+\left(\dfrac{16600939332793}{1098809712000} - \dfrac{12568}{3465}{\rm eulerlog}_{4}(x)+\nu c_{44}^{\rm 3PN}(\nu)\right)x^{3},\\
\rho_{43}(x;\nu)&=\rho_{43}^{\rm 1PN}(x:\nu)-\left(\dfrac{6894273}{7047040}+\nu c_{43}^{\rm 2PN}(\nu)\right)x^{2},\\
\rho_{42}(x;\nu)&=\rho_{42}^{\rm 2PN}(x;\nu)+\left(\dfrac{848238724511}{219761942400} - \dfrac{199276197120}{219761942400}{\rm eulerlog}_{2}(x)+\nu c_{42}^{\rm 3PN}(\nu)\right)x^{3}\\
\rho_{41}(x;\nu)&=\rho_{41}^{\rm 1PN}(x;\nu)-\left(\dfrac{7775491}{21141120}+\nu c_{41}^{\rm 2PN}(\nu)\right)x^{2},\\
\rho_{55}(x;\nu)&=\rho_{55}^{\rm 1PN}(x;\nu)-\left(\dfrac{3353747}{2129400}+\nu c_{55}^{\rm 2PN}(\nu)\right)x^{2},\\
\rho_{53}(x;\nu)&=\rho_{53}^{\rm 1PN}(x;\nu)-\left(\dfrac{410833}{709800}+\nu c_{53}^{\rm 2PN}(\nu)\right)x^{2},\\
\rho_{51}(x:\nu)&=\rho_{51}^{\rm 1PN}(x;\nu)-\left(\dfrac{31877}{304200}+\nu c_{51}^{\rm 2PN}(\nu)\right)x^{2}.
\end{align}
\end{widetext}
To go up to 5.5PN order we also retain all multipoles up to $\ell=6$, but only the Newtonian or 1PN $\nu$-dependence is important,
so that it is not needed to indicate higher coefficients.
Then, by expanding Eq.~\eqref{eq:flux} up to 5.5PN and defining
the Newton-normalized flux as
$\hat{\cal F}\equiv {\cal F}/F_{22}^{\rm Newt}(x)$, we finally
obtain the following terms beyond 3.5~PN and up to 5.5~PN:
\begin{widetext}
\begin{align}
\hat{\cal F}_{\rm 4PN}&=\bigg\{\dfrac{\nu}{9}\left(1-4\nu \right) c_{21}^{\rm 4PN}(\nu)+\nu  \left(4c_{22}^{\rm 4PN}(\nu)-\frac{5429939 \pi ^2}{96768}+\frac{196922 \gamma_{\rm E}
   }{2205}-\frac{2580408938233367}{3203802201600}\right)+\frac{4439809795 \nu ^4}{14239120896}\nonumber\\
   &+\frac{20946233683 \nu^3}{1456273728}+\left(\frac{54048871600249}{320380220160}-\frac{136735 \pi ^2}{24192}\right) \nu ^2+\left(\frac{47385 \nu
   }{392}-\frac{47385}{1568}\right) \log (3)\nonumber\\
   &+\left(\frac{126302 \nu }{2205}+\frac{39931}{294}\right) \log (2)+\left(\frac{98461 \nu
   }{2205}+\frac{232597}{8820}\right) \log (x)-\frac{1369 \pi ^2}{126}+\frac{232597 \gamma_{\rm E} }{4410}-\frac{323105549467}{3178375200}\bigg\}x^4,
\end{align}
\begin{align}
  \hat{\cal{F}}_{\rm 4.5 PN}&=\pi \Bigg( -\frac{3719141 \nu ^3}{38016}-\frac{133112905 \nu ^2}{290304}
  +\left(\frac{2062241}{22176}+\frac{41}{12}\pi^2\right)\nu
  +\frac{265978667519}{745113600}-\frac{6848}{105}{\rm eulerlog}_2(x)\Bigg)x^{9/2},
\end{align}
\begin{align}
  \label{eq:hF5PN}
\hat{\cal{F}}_{\rm 5PN}&=\Bigg\{-\dfrac{1}{1-3\nu}
\Bigg[\dfrac{2500861660823683}{2831932303200}
- \dfrac{10608155067013217101261}{1904961566258150400}\nu
+\dfrac{24823886833955583083459}{3265648399299686400}\nu^2\nonumber\\
&+\dfrac{83916740839405609479263}{68578616385293414400}\nu^3
+ \dfrac{31282558612344304720193}{3809923132516300800}\nu^4
- \dfrac{163716515443482656797}{35168521223227392}\nu^5\nonumber\\
&+ \dfrac{1247153383060759670851}{1714465409632335360}\nu^6
+\gamma_{\rm E}\left(-\dfrac{916628467}{7858620} + \dfrac{3232420865}{3667356}\nu
- \dfrac{12533953163}{9168390}\nu^2 - \dfrac{2084604421}{3056130}\nu^3\right)\nonumber\\
&+\pi^2\left(\dfrac{424223}{6804} - \dfrac{520672457}{1354752}\nu
+ \dfrac{1244952467}{4064256}\nu^2 + \dfrac{888697001}{1016064}\nu^3
- \dfrac{36678805}{677376}\nu^4\right)\nonumber\\
&+\left(\dfrac{83217611}{1122660} - \dfrac{11385383927}{18336780}\nu
+ \dfrac{12117535706}{4584195}\nu^2 - \dfrac{1327348993}{305613}\nu^3\right)\log(2)\nonumber\\
&+\left(-\dfrac{47385}{196} + \dfrac{1405755}{784}\nu - \dfrac{2827305}{784}\nu^2 + \dfrac{236925}{196}\nu^3\right)\log(3)\nonumber\\
&+\left(-\dfrac{916628467}{15717240}
+ \dfrac{3232420865}{7334712}\nu
- \dfrac{12533953163}{18336780}\nu^2
- \dfrac{2084604421}{6112260}\nu^3\right)\log(x)\Bigg]\nonumber\\
&+\left(-\dfrac{114}{7}+\dfrac{55}{7}\nu\right)\nu c_{22}^{\rm 4PN}(\nu) +4\nu c_{22}^{\rm 5PN}(\nu) +\frac{1}{9}\nu(1-4\nu)c_{21}^{\rm 4PN}+\left(-\dfrac{1}{56}+\dfrac{151}{756}\nu-\dfrac{97}{189}\nu^{2}\right)\nu c_{21}^{\rm 3PN}(\nu) \nonumber\\
   &+\nu(1-4\nu)\left(\dfrac{1}{1344}c_{31}^{\rm 4PN}(\nu)+\dfrac{3645}{448}c_{33}^{\rm 4PN}(\nu)\right)
     +\nu(1-3\nu)^{2}\left(\dfrac{10}{21}c_{32}^{\rm 3PN}(\nu)+\dfrac{10240}{567}c_{44}^{\rm 3PN}(\nu)+\dfrac{40}{3969}c_{42}^{\rm 3PN}(\nu)\right)\nonumber\\
   &-\nu(1-2\nu)^{2}(4\nu-1)\Bigg(\dfrac{729}{560}c_{43}^{\rm 2PN}+\dfrac{1}{35280}c_{41}^{{\rm 2PN}}(\nu)+\dfrac{48828125}{1216512}c_{55}^{\rm 2PN}(\nu)
   +\dfrac{2187}{45056}c_{53}^{\rm 2PN}(\nu)+\dfrac{1}{12773376}c_{51}^{\rm 2PN}(\nu)\Bigg)
\Bigg\}x^{5}
\end{align}
\begin{align}
  \label{eq:hF55PN}
\hat{\cal F}_{\rm 5.5PN}&=
\pi \Bigg\{ \dfrac{2\nu}{9}\left(1-4\nu\right) c_{21}^{\rm 3PN}(\nu)
+\frac{6058253667029 \nu^4}{64076044032}+\frac{70029960211823 \nu^3}{85434725376}+\frac{100004171503889 \nu^2}{160190110080}\nonumber\\
&+\nu  \left(16c_{22}^{\rm 4PN}(\nu)-\frac{138140205552539713}{25630417612800}\right)+\frac{8399309750401}{101708006400}
  +\pi^2 \left(\frac{1015201 \nu }{193536}-\frac{1605601 \nu^2}{48384}\right)\nonumber\\
  &+\gamma_{\rm E}  \left(\frac{2781341 \nu }{4410}+\frac{177293}{1176}\right)+\left(\frac{2360189 \nu }{4410}+\frac{8521283}{17640}\right) \log (2)+ \left(\frac{142155 \nu
   }{196}-\frac{142155}{784}\right) \log (3) \nonumber\\
   &+ \left(\frac{2781341 \nu }{8820}+\frac{177293}{2352}\right) \log (x)\Bigg\}x^{11/2}.
\end{align}
\end{widetext}
A few comments are needed. First, even if the 4PN calculation of the flux is not fully completed,
thanks to the fact that both energy and angular momentum are actually known at 4PN,
one finds that the 4PN total flux only depends on {\it two} functions of $\nu$ that parametrize
the missing 4PN $\nu$ dependence in the $\ell=2$ amplitude waveform multipoles. 
Interestingly, once both $c_{22}^{\rm 4PN}(\nu)$ and $c_{21}^{\rm 4PN}(\nu)$ will be known,
because of the resummed structure of the tail factor, one will also know the 5.5PN tail term,
that only depends on 4PN information coming from the energy, angular momentum and residual waveform amplitudes,
likewise the 4.5PN tail term, that only depends on 3PN information with the same origin.
In fact, we see that the 4.5PN flux term is free of uknown coefficients and coincides
precisely, seen the definition of the ${\rm eulerlog}_2(x)$ function,
with Eq.~(5.10) of Ref.~\cite{Marchand:2016vox} obtained from first principles using the MPM approach.
Note also that, even in the absence of full analytical knowledge of the various $c_\lm^{\rm nPN}$,
one may think to extract these parameters from long-inspiral, radiation-reaction dominated,
NR waveforms, using the parametrized $\rho_{\ell m}$'s written above within the EOB model.
We will give a flavor of this idea in Sec.~\ref{sec:F2nonspinning} below, though in a
simplified scenario that uses the IMRPhenomD~\cite{Khan:2015jqa} phenomenological waveform
model as an approximation to the NR waveform, and the (parametrized) 4.5PN-accurate Taylor~F2
description of the Fourier-domain phase (in the stationary phase approximation)
instead of the time-domain EOB one.

\section{NLO spin-orbit and LO spin-spin tail induced flux terms}
\label{sec:FtailNLO}
The same procedure outlined above can be applied, in the case of spin, to show that also the next-to-leading tail-induced
spin-orbit term in the flux obtained by Marsat et al.~\cite{Marsat:2013caa} using the MPM formalism 
is contained in Eqs.~\eqref{eq:flux}-\eqref{eq:hF} above once the spin-dependent information is suitably included   
in the various functions. We will have to work at 4PN fractional order, i.e. retain in the expansion of $|T_{\ell m}|^{2}$ 
terms up to $x^{4}$. To push the calculation up to this order, we need the spin-orbit NLO $\rho_{\lm}^{S}$ for multipoles 
with $l=2$ (in particular, the $\tilde f_{\lm}$ factorization from~\cite{Damour:2014sva} was used for the odd one), 
whereas for the multipoles with $\ell=3$ and $\ell=4$ the spin relativistic amplitude corrections have been utilized at 
the leading order (here the $\tilde f_{\lm}$ were used for the (3,3), (3,1), (4,3) and (4,1) multipoles). 
We do so by using the NLO terms of Ref.~\cite{Damour:2014sva} and the LO terms of Ref.~\cite{Pan:2010hz}, but,
instead of expressing them, as it is customary, by means of the usual symmetric and antisymmetric combinations 
$\chi_{S}\equiv (\chi_{1}+\chi_{2})/2$ and $\chi_{A}\equiv (\chi_{1}-\chi_{2})/2$, of the dimensionless
spins $\chi_{1,2}\equiv S_{1,2}/(m_{1,2})^{2}$, we use instead
\be
\tilde{a}_{i}\equiv X_{i} \chi_{i}\qquad i=1,2\,,
\ee
where $X_{i}\equiv m_{i}/M$. In the usual convention that $X_{1}\geq X_{2}$, that is $X_{1}=1/2\left(1+\sqrt{1-4\nu}\right)$
and $X_{2}=1-X_{1}$. This choice is convenient for two reasons: (i) the analytical expression get more compact as several
factors $\sqrt{1-4\nu}$ are absorbed in the definitions,
and one can clearly distinguish the sequence of terms that are ``even'', in the sense that are symmetric under exchange of body
1 with body 2 and are proportional to the ``effective spin'' $\hat{a}_{0}\equiv \tilde{a}_{1}+\tilde{a}_{2}$
from those that are ``odd'', i.e. change sign under the exchange of body 1 with body 2 and are proportional to the factor 
$\sqrt{1-4\nu}(\tilde{a}_{1}-\tilde{a}_{2})$; (ii) in addition, one can see the test-particle limit just inspecting visually the 
formulas, since in this limit, $m_{2}\ll m_{1}$, $\tilde{a}_{2}\to 0$ and $\tilde{a}_{1}$ becomes the dimensionalmente spin of 
the massive black hole of mass $m_{1}=M$. Similarly, one can recover the {\it spinning} particle limit because in that
limit $\tilde{a}_{2}$ just reduces to the usual spin-variable of the particle $\sigma \equiv S_{2}/({m_{1} m_{2}})$.
Moreover, the orbital $\rho_{\lm}$s were taken at 3PN for the (2,2) multipole; at 2PN for (2,1), (3,2),(3,1),(3,3), (4,4) and (4,2) 
multipoles and at 1PN for (4,1) and (4,3) multipoles.   All these truncations are consistent with the factorizations of the 
Newtonian prefactors. Beyond that, one needs to gather all flux multipoles up to $\ell=m=6$ and only the even parity multipoles $(7,7)$, $(7,5)$, $(7,3)$ and $(7,1)$ for $\ell=7$, since the other multipoles enter at  higher PN order. 
More precisely, for $(7,m)$ and $(6,m)$ with $m={\rm odd}$ only the Newtonian contribution is needed; 
for $(6,m)$ and $(5,m)$ with $m={\rm even}$ one needs only the orbital part up to 1PN; for $(5,m)$ with $m={\rm odd}$ one needs only the orbital part up to 2PN.
Using this spin-dependent information in Eqs.~\eqref{eq:flux}-\eqref{eq:hF}, a straightforward calculation gives
\begin{widetext}
\be
\label{eq:4PNtail}
\hat{\cal F}^{\rm SO}_{\rm 4PN}=-\pi\left[(\tilde{a}_{1}+\tilde{a}_{2})\left[\frac{3485}{192}+\dfrac{13879}{144}\nu\right]-
\sqrt{1-4\nu}\,(\tilde{a}_{1}-\tilde{a}_{2})\left(\dfrac{10069}{1344}+\dfrac{21241}{672}\nu\right)\right]x^{4},
\ee
\end{widetext}
which coincides with the NLO term of Eq.~(4.9) of Ref.~\cite{Marsat:2013caa} once written using $\tilde{a}_{1,2}$.
In doing so, one also notices that, by also including the leading-order spin-spin term in $\rho_{22}^{\rm S}$ as computed in
Ref.~\cite{Damour:2014sva} (see Eq.~(81) and~(87) there)
\be
\rho_{22}^{\rm SS_{{\rm LO}}} = \dfrac{1}{2}\left(\tilde{a}_{1}+\tilde{a}_{2}\right)x^{2},
\ee
one can immediately obtain the leading-order, tail-induced, spin-spin term in the flux, as
\be
\label{eq:ssTail}
\hat{{\cal F}}^{\rm SS_{{\rm LO}}}_{\rm 3.5PN}=\pi\left(\dfrac{65}{8}\tilde{a}_{1}^{2}+\dfrac{63}{4}\tilde{a}_{1}\tilde{a}_{2}+\dfrac{65}{8}\tilde{a}^{2}\right)x^{7/2}.
\ee
In the test-particle limit, $\tilde{a}_{2}=0$, this term is just the well-known test-particle limit in the flux
obtained in Eq.~(G19) of Ref.~\cite{Tagoshi:1996gh}. To our knowledge, this leading-order tail-induced
spin-spin term is obtained here  for the first time.
Evidently, the procedure can be extended to higher orders and one can obtain higher PN terms in 
the flux, either even, or odd, that depend on coefficients that are given a clear physical meaning
within the EOB formalism. However, let us note that, differently from the nonspinning case, one {\it is not} 
allowed to introduce in the $\rho_{\lm}$'s terms that are obtained in the test-particle limit and parametrize 
the rest, as it is done for the orbital part. The reason is that the test-particle results
come from the combination of two kinds of terms, some proportional to $(\tilde{a}_{1}+\tilde{a}_{2})$
and some other proportional to $\sqrt{1-4\nu}(\tilde{a}_{1}-\tilde{a}_{2})$, likewise what is
seen in Eq.~\eqref{eq:4PNtail} above. 

\section{4.5PN parametrized Taylor~F2 approximant: the nonspinning case}
\label{sec:F2nonspinning}
\begin{figure}[t]
\center
\includegraphics[width=0.45\textwidth]{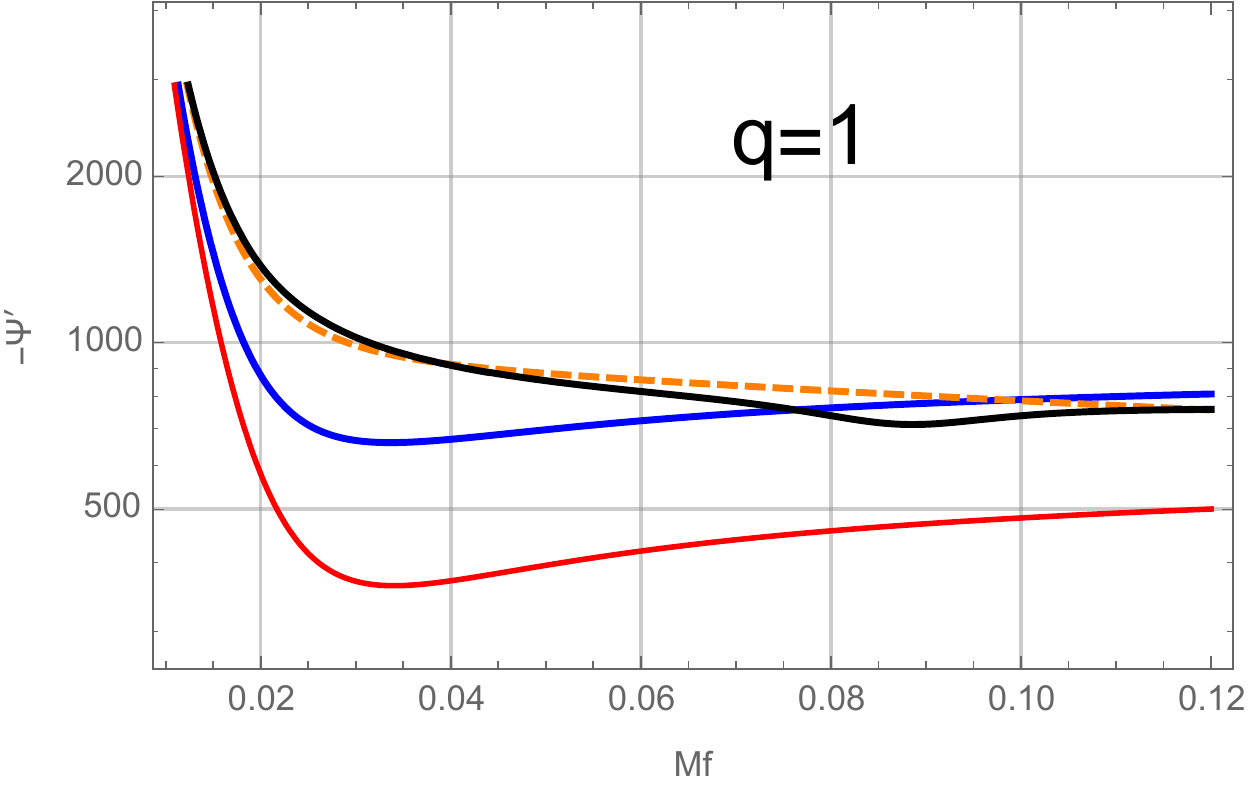}
\caption{\label{fig:q1Dpsi0}Nonspinning case, $q=1$. Comparison betweeen different
  representations of the derivative of the phase of the Fourier transform, $-\Psi'(f)$.
  The plot shows: (i) the standard 3.5PN Taylor~F2 approximant (blue); (ii) the parametrized 4.5PN one
  with $c_{22}^{\rm 4PN}=0$ (red); the {\it tuned} parametrized 4.5PN one with $c_{22}^{\rm 4PN}=-200$ (orange, dashed);
  (iii) the phase  derivative of the IMRPhenomD model~\cite{Khan:2015jqa} (black). A single parameter is
  able to improve the consistency between $T_{\rm F2}^{\rm 4.5PN}$ and IMRPhenomD even
  below the standard Schwarzschild LSO frequency $Mf_{\rm LSO}^{\rm Schw}\approx 0.0217$.}
\end{figure}
\begin{figure*}[t]
  \center
  \includegraphics[width=0.45\textwidth]{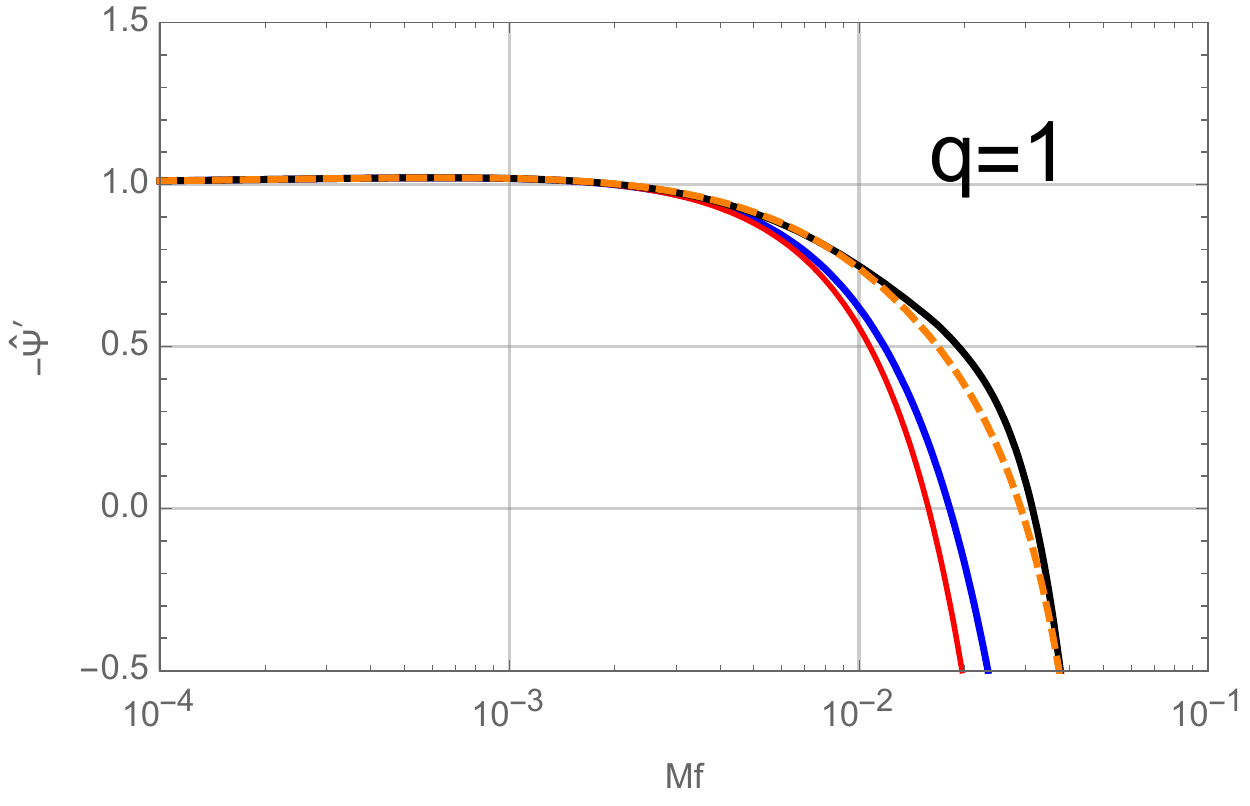}
  \includegraphics[width=0.45\textwidth]{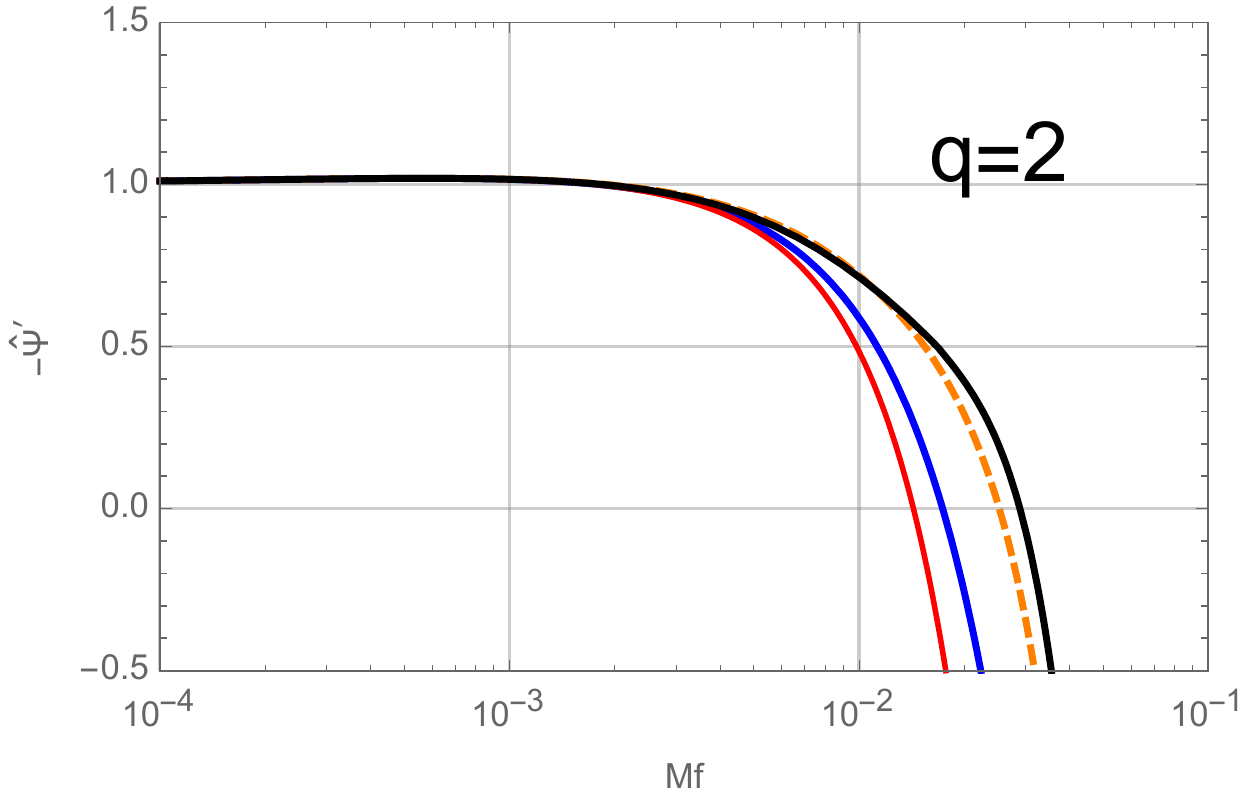}\\
  \includegraphics[width=0.45\textwidth]{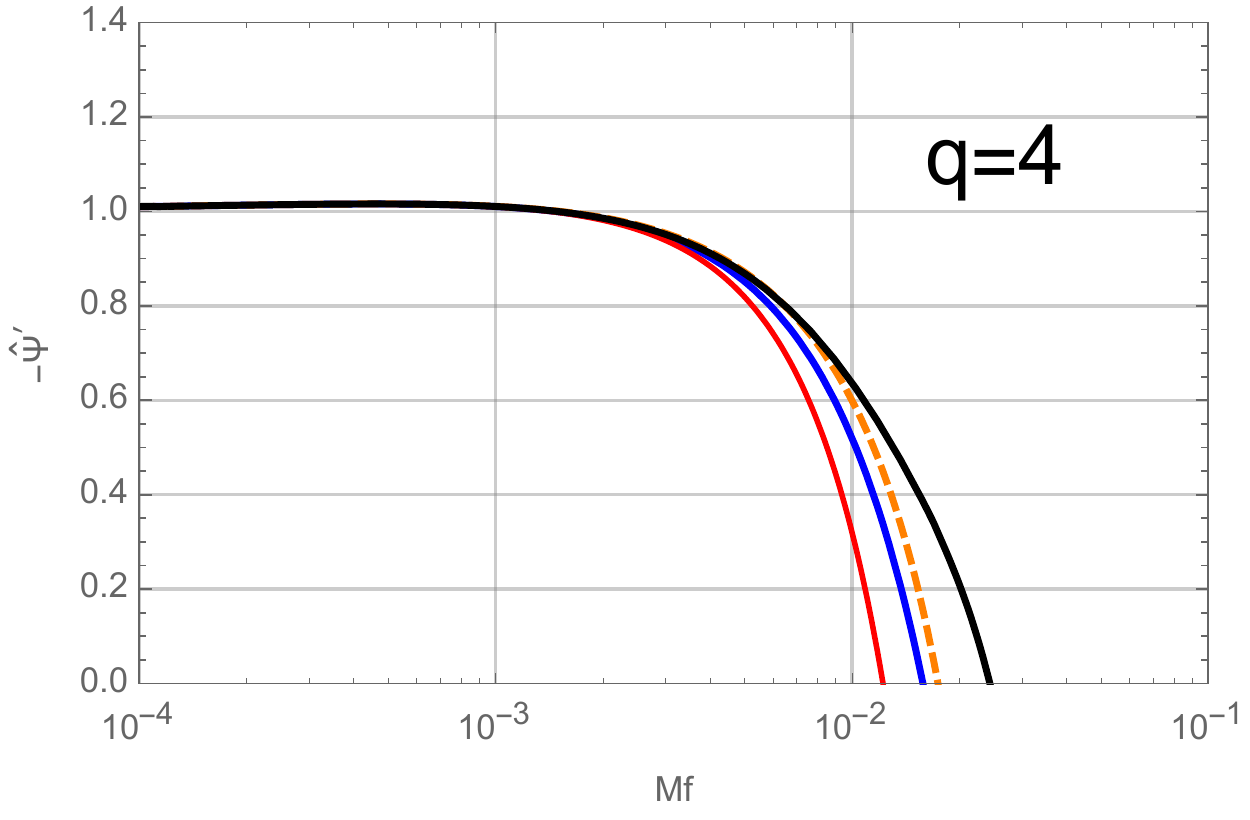}
  \includegraphics[width=0.45\textwidth]{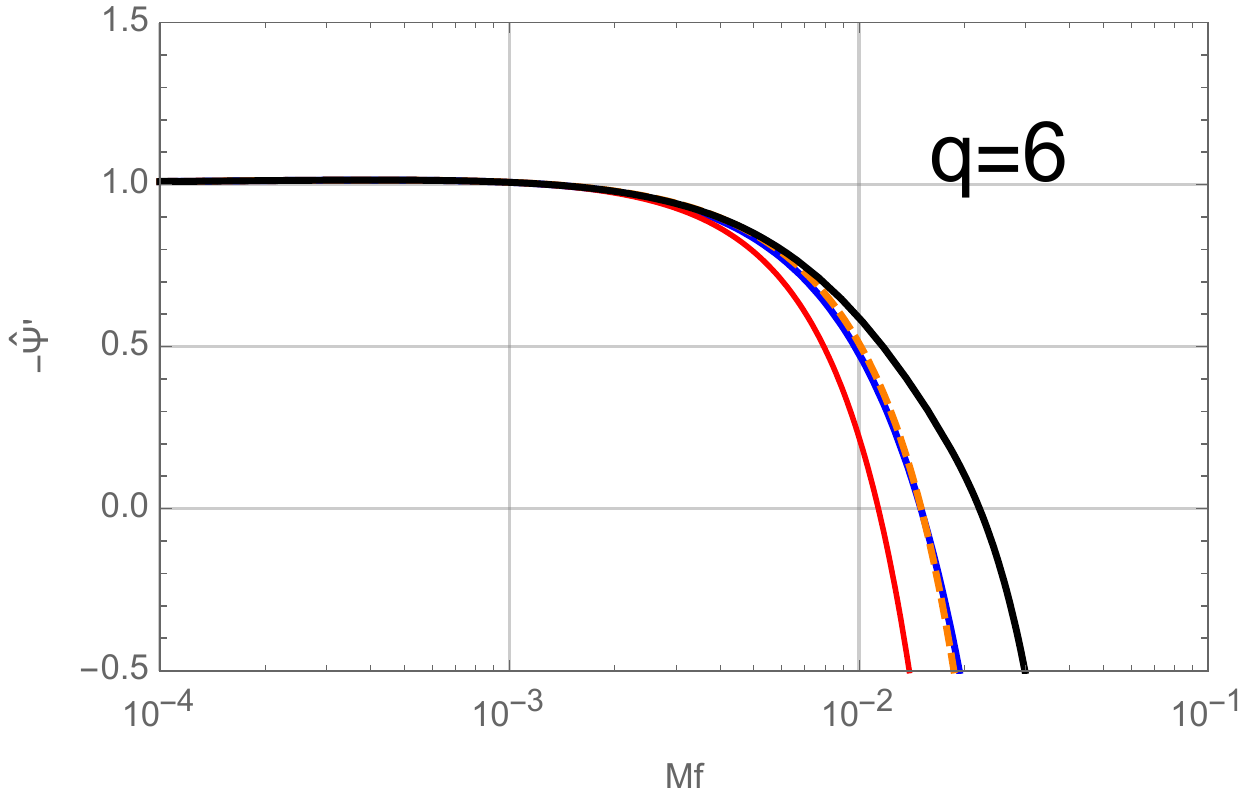}\\
  \includegraphics[width=0.45\textwidth]{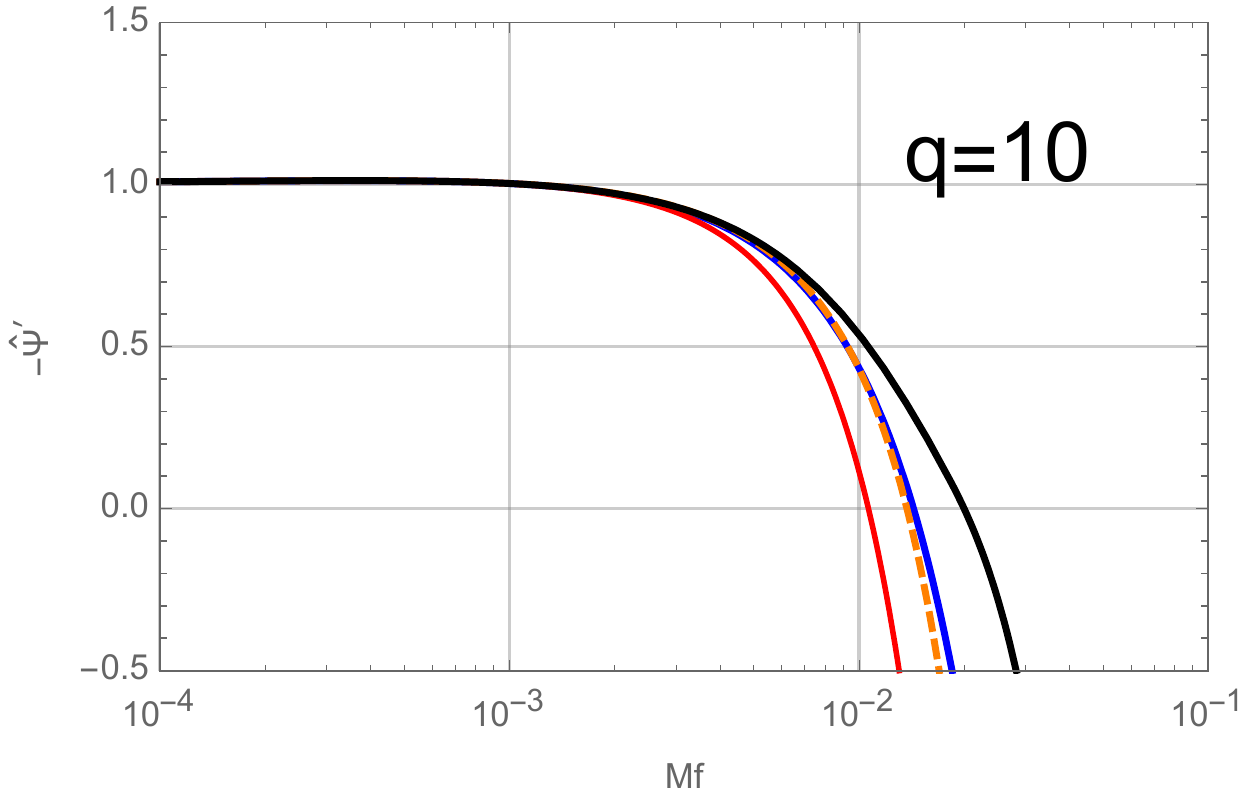}
    \includegraphics[width=0.45\textwidth]{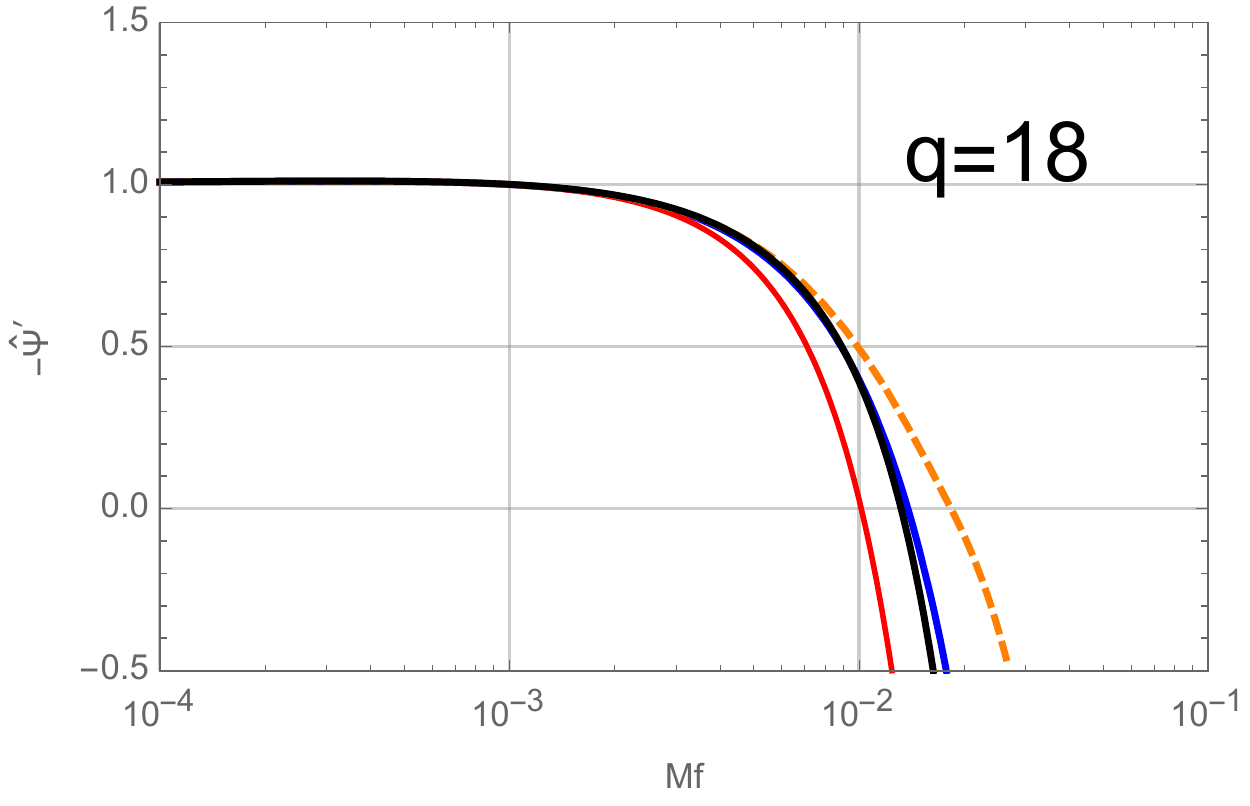}
    \caption{\label{fig:hatDpsi}Qualitative exploration of the flexibility of the parametrized,
      nonspinning, 4.5PN Taylor F2 approximant. The figure shows the Newton-normalized derivative
      of the frequency-domain phase for mass ratios $q=(1,2,4,6,10,18)$, Eq.~\eqref{eq:hatPsi},
      computed for: (i) the IMRPhenomD phase (black); (ii) the 3.5PN approximant (blue); (iii)
      the 4.5PN approximant without tuning (red), $c_{22}^{\rm 4PN}=c_{21}^{\rm 4PN}=0$;
      (iv) the 4.5PN approximant with $c_{21}^{\rm4PN}=0$ and $c_{22}^{\rm 4PN}$ tuned so to obtain
      an accurate representation the IMRPhenomD phase on the largest possible frequency interval.}
\end{figure*}
As a straightforward application of our calculation of the flux
at (parametrized) 4.5PN we can compute the corresponding
Taylor~F2 frequency domain approximant, using the SPA,
to the phase of the Fourier transform of the signal.
Here, as an exploratory, proof-of-principle, investigation,
we only focus on the phase of the Fourier transform and do
not discuss the amplitude. The TaylorF2 frequency-domain
waveform phase up to 4.5PN reads
\begin{align}
\label{eq:F2}
\Psi_{4.5}^{\rm F2}(f)&=2\pi f t_c - \varphi_c-\dfrac{\pi}{4}\nonumber\\
                    &+\dfrac{3}{128\nu}(\pi fM)^{-5/3}\sum_{i=0}^{9}\varphi_i(\pi fM)^{i/3},
\end{align}
where the $\varphi_i$'s with $i={0,1,\dots,7}$ are given by the nonspinning limit of Eqs.~(B6)-(B13)
Ref.~\cite{Khan:2015jqa}. The additional 4PN (parametrized) and 4.5PN (exact) terms read
\begin{widetext}
  \begin{align}
\label{eq:phi8}
   \varphi_8&= [1-\log (\pi M f)] \biggl[c_{21}^{\rm 3PN} \biggl(\frac{40 \nu }{81}-\frac{160 \nu ^2}{81}\biggr)+\frac{160 c_{22}^{\rm 4PN} \nu
   }{9}-\frac{369469478275 \nu ^4}{16019011008}+\frac{510041481025 \nu
   ^3}{13106463552}\\\nonumber
   &+\biggl(\frac{300600673165997}{576684396288}-\frac{399545 \pi ^2}{27216}\biggr) \nu ^2+
   \biggl(-\frac{5679872289503527}{1281520880640}-\frac{5322928 \gamma }{3969}+\frac{9302215 \pi ^2}{54432}\\\nonumber
   &-\frac{1420688 \log
   (2)}{441}+\frac{26325 \log (3)}{49}\biggr)\nu-\frac{90490 \pi ^2}{567}-\frac{36812 \gamma
   }{189}+\frac{2550713843998885153}{830425530654720}-\frac{26325 \log (3)}{196}\\\nonumber
   &-\frac{1011020 \log
   (2)}{3969}\biggr]+\biggl(\frac{2661464 \nu }{11907}+\frac{18406}{567}\biggr)\log^2 (\pi M f),\\
  \label{eq:phi9}
  \varphi_{9}&=\pi\bigg(-\frac{13696}{63}   \log (\pi M f)+\frac{10323755  \nu^3}{199584}+\frac{45293335}{127008}\nu^2
  +\left(\frac{2255 \pi^2}{6}-\frac{1492917260735}{134120448}\right)\nu
  -\frac{640 \pi^2}{3}-\dfrac{13696}{21} \gamma_{\rm E} \nonumber\\
           &+\frac{105344279473163}{18776862720}-\frac{27392}{21}  \log (2)\bigg).
\end{align}
\end{widetext}
\begin{figure*}[t]
  \center
  \includegraphics[width=0.45\textwidth]{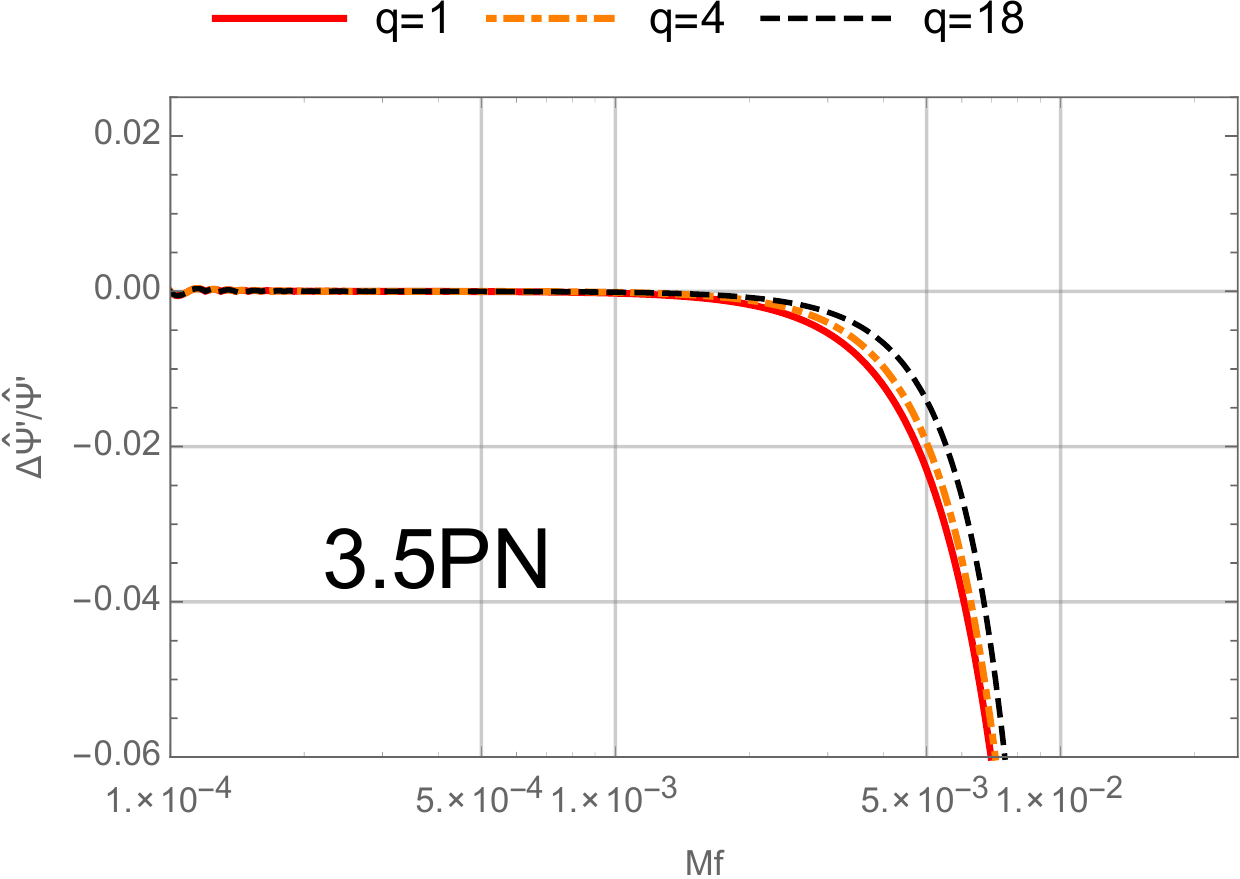}
  \includegraphics[width=0.45\textwidth]{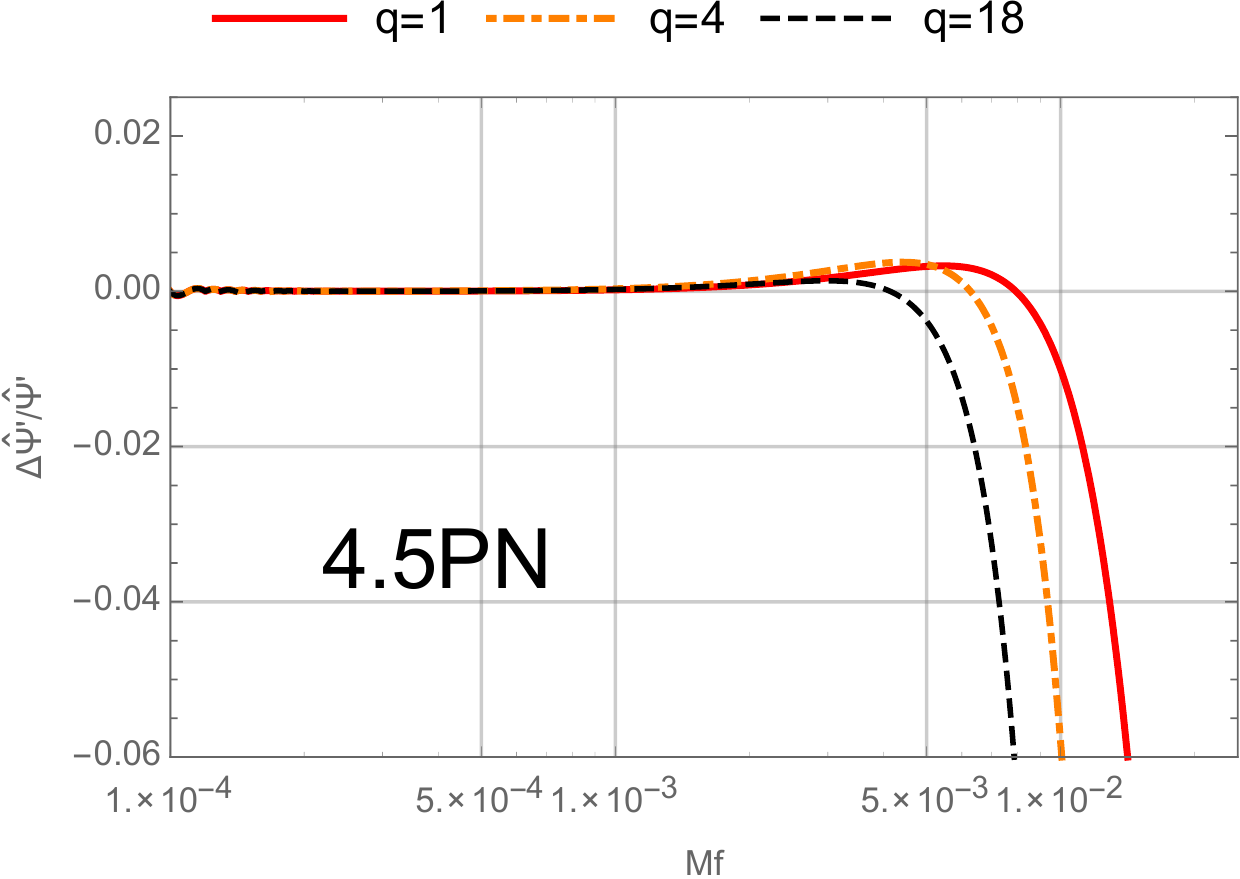}
  \caption{\label{fig:RelDiff}Relative differences between IMRPhenomD
    and TaylorF2 PN approximants: the 3.5PN disagreement is
    reduced with the 4.5PN tuned approximant, especially for
    nearly equal-mass binaries.}
\end{figure*}
In this way we have constructed an approximant that, though dependent on some unknown
analytical information, it incorporates in fully consistent way 4PN and 4.5PN terms
with the complete $\nu$ dependence. The residual ignorance only resides in
the functions $c_{22}^{\rm 4PN}(\nu)$ and $c_{21}^{\rm 4PN}(\nu)$ that 
parametrize currently unknown 4PN waveform terms. 
A priori, one is expecting these functions to be just
third-order polynomials in $\nu$ by induction from the previous PN orders, since
the highest power of $\nu$ that can appear coincides with the highest PN order considered
(recall that the quantities that appear in the fluxes are $\nu c_{\ell m}^{\rm 4PN}(\nu)$).
Note that, even before the 4PN waveform calculation will be completed, it would be in principle possible
to constrain $(c_{22}^{\rm 4PN}(\nu),c_{21}^{\rm 4PN}(\nu))$ by comparing {\it parametrized}
EOB predictions with NR waveform data, a line of research that worths being investigated
in detail in the future. More modestly, here we try to illustrate the usefulness
of the new terms in Eqs.~\eqref{eq:phi8}-\eqref{eq:phi9} by presenting 
a simple comparison between: (i) the new 4.5PN-accurate, parametrized, TaylorF2 approximant;
(ii) the usual 3.5PN accurate approximant; (iii) the IMRPhenomD phenomenological model of Fourier
phasing of Ref.~\cite{Khan:2015jqa}. This model is calibrated to hybrid EOB-NR waveforms and is
able to yield an accurate representation of our best description of the actual phasing
(for spinning, nonprecessing, BBHs) from inspiral to plunge, merger and ringdown
on a large portion  of the (nonprecessing) BBH parameter space. 
During the inspiral (more precisely, up to $Mf=0.018$),
it is built by {\it flexing} the standard 3.5PN Taylor~F2 approximant by adding
effective, parameter-dependent, high-order PN terms that complement
the 3.5PN analytical knowledge for both the Fourier phase and amplitude.
The terms that are added to do so are effectively some $(\varphi_{8},\varphi_{9},\varphi_{10},\varphi_{11})$.
The difference with our case is that no partial analytical knowledge is assumed 
on them; rather, these terms are determined by fitting to the Fourier transform
of suitably computed EOB-NR hybrid waveforms. 
The IMRPhenomD model was proved to be faithful against a large portion 
of the currently available Simulating eXtreme Spacetimes (SXS) catalog~\cite{SXS:catalog} 
of NR waveforms; for this reason, in this work we assume it is delivering the ``exact''
representation of the Fourier phasing (certainly for $Mf\geq 0.0035$, that is
the minimum frequency considered for doing the phenomenological fits,
see Sec.~VIA of~\cite{Khan:2015jqa}), and thus we use it as a benchmark for
the various PN approximant.
We compare the {\it derivative} of the phase of the Fourier transform with
respect to the frequency, $-\Psi'(f)\equiv -\de \Psi(f)/\de f$, since this is
the quantity that was directly calibrated to the hybrid waveforms in Ref.~\cite{Khan:2015jqa}
for technical reasons. For simplicity, we do so by putting $t_{c}=0$ both in
the definition of $\Psi_{4.5}^{\rm F2}$, Eq.~\eqref{eq:F2} above as well
as in the IMRPhenomD model. Figure~\ref{fig:q1Dpsi0} contrasts four
representations of $-\Psi'(f)$ for the equal-mass, $q=1$
case~\footnote{Note however that for this figure the arbitrary
  gauge parameter $t_c$ is taken different from zero, but the
  same for all curves, just for visualization purposes.}
The black line is the IMRPhenomD function; the blue is the standard 3.5PN TaylorF2
approximant, while the red is the 4.5PN approximant with $c_{22}^{\rm 4PN}=0$.
Not surprisingly, this latter is neither better nor worse than the 3.5PN.
However, we find that by considering $c_{22}^{\rm 4PN}$
as an {\it effective}, tunable, parameter it is fairly easy to lower the red curve
in the figure and to move it on top of the black one during the inspiral.
Interestingly, the agreement remains good also {\it after} $fM\approx 0.02$.
We remind, as a useful mnemonic reference point, that the GW frequency of
the Schwarzschild last-stable-orbit (LSO) is $(Mf)_{\rm LSO}^{\rm GW}=\pi^{-1} 6^{-3/2}\approx 0.0217$.
Note however that for an equal-mass, nonspinning system the LSO, as defined
within the EOB approach, occurs at higher frequencies because of the $\nu$-dependent
terms in the interaction potential~\cite{Buonanno:1998gg}.
The orange-dashed curve in Fig.~\ref{fig:q1Dpsi0} is obtained by fixing
$c_{22}^{\rm 4PN}=-200$ (note in this respect that for $\nu=1/4$
the $c_{21}^{\rm 4PN}$-dependent term just vanishes in Eq.~\eqref{eq:phi8}).
One then finds that the parametrized 4.5PN approximant is sufficiently
robust to allow one to obtain similar results also for larger values of $q$:
keeping $c_{21}^{\rm 4PN}=0$ for simplicity, it is always possible to find
a value of $c_{22}^{\rm 4PN}$ that reduces the gap between the 4.5PN and
the IMRPhenomD curves in a frequencies interval where the 3.5PN approximant
is not reliable anymore. To quantify this effect, it is useful to look at
the Newton-normalized derivative of the phase
\be
\label{eq:hatPsi}
\hat{\Psi}'(f)=\Psi'(f)\left(\dfrac{5}{128 (f M)^{8/3}\pi^{5/3}\nu}\right)^{-1},
\ee
In Fig.~\ref{fig:hatDpsi} we show this quantity for six values of the mass ratio $q$
that belong to the domain of calibration of IMRPhenomD~\cite{Khan:2015jqa}.
We use as lower frequency limit $fM=0.0001$. As in Fig.~\ref{fig:q1Dpsi0},
the figure compares four curves: the IMRPhenomD (black); the TaylorF2 at 3.5PN (blue);
the {\it untuned} TaylorF2 at 4.5PN  ($c_{22}^{\rm 4PN}=c_{21}^{\rm 4PN}=0$);
the $c_{22}^{\rm 4PN}$-tuned TaylorF2 at 4.5PN. The parameter is determined
by hand so to find good consistency with the black curve on the largest
possible frequency interval: this is done by inspecting the (relative)
difference with $\psi'(f)$ yielded by the IMRPhenomD model
and tuning $c_{22}^{\rm 4PN}$ so that this difference remains flat on
the largest possible frequency interval (see below). Figure~\ref{fig:hatDpsi}
illustrates several points: (i) the low-frequency consistency between
all PN approximants as well as IMRPhenomD; (ii) neither the 3.5PN-accurate
nor the untuned 4.5PN-accurate TaylorF2 deliver accurate approximations
to the IMRPhenomD inspiral phasing after frequency $fM\approx 0.005$;
(iii) by tuning {\it only} the parameter $c_{22}^{\rm 4PN}$ one can easily
put the 4.5PN curves on top of the IMRPhenomD for all configurations and
the good agreement is kept essentially up to $fM\approx 0.01$ for all
binaries (or even to $fM\approx 0.2$ for $q\simeq 1$) except for $q=18$,
where the consistency is trustable up to $fM\approx 0.005$.
The good values of $c_{22}^{\rm 4PN}(q)$ yielding such behaviors
for $q=(1,2,3,4,6,7,10,18)$ are $c_{22}^{\rm 4PN}=(-200, -290,-390,-480, -650, -750,-1050,-2000)$.
The improvement brought by the 4.5PN tuned approximant with respect to
the purely analytical 3.5PN one is quantitatively illustrated in Fig.~\ref{fig:RelDiff},
that shows the {\it relative differences} with the IMRPhenomD, i.e.
$\Delta\Psi'/\Psi' \equiv (\Psi'_{\rm IRMPhenomD}-\Psi'_{\rm *PN})/\Psi'_{\rm IMRPhenomD}$ and
highlights how the region of frequencies reliably covered by the PN
approximant is enlarged when the 4.5PN-tuned description is employed.
Finally, it is easily found that the ``first-guess'' values of
$c_{22}^{\rm 4PN}$ mentioned above and used in Figs.~\ref{fig:hatDpsi}-\ref{fig:RelDiff}
can be well fitted by a rational function of the form
\be
\label{eq:c4PN}
c_{22}^{\rm 4PN}(\nu)=c_{1/4}^{\rm 4PN}\dfrac{1+n_{1}(X_{12})^{2} + n_{2}(X_{12})^{4}+n_{3}(X_{12})^{6}}{1+d_1(X_{12})^{2}}
\ee
where $X_{12}\equiv X_{1}-X_{2}=\sqrt{1-4\nu}$, with $c_{1/4}^{\rm 4PN}=-200.821$, $n_{1}=3.1334$, $n_{2}=-6.4605$, $n_{3}=5.0608$, $d_{1}=-1.0027$.

In principle, this effective representation for $c_{22}^{\rm 4PN}$ can yield a
complete 4.5PN TaylorF2 approximant that could replace the purely
analytical 3.5PN one for the nonspinning sector. Still, we would like to remain
very conservative at this stage and stress that our analysis should be
considered mainly qualitative and only  marginally quantitative, since
it only aims at showing the potentialities of our approach.
A more quantitative analysis is beyond the scope of this
work, and it should be done, for example, by comparing the
parametrized PN approximant with long-inspiral EOB waveforms,
in order to further reduce potential uncertainties in
the IMRPhenomD model. In this respect, a question that may be interesting
to answer is to which accuracy an EOB-tuned 4.5PN approximant can effectively
reproduce the EOBNR phasing and up to which value of the total mass (probably higher
than the limits given by the 3.5PN one~\cite{Buonanno:2009zt}) of the binary it could
be used in searches. These questions will hopefully find their answers in future studies.

In addition, we also would like to note that the parametrized 4.5PN
(once extended to the spinning case) approximant might possibly
be used as a replacement of the phenomenological ansatz given by
Eq.~(28) of Ref.~\cite{Khan:2015jqa}, so to include more analytic
information and possibly simplify the fitting procedures, with
less parameters to be determined by fitting. Evidently, in doing
so one may investigate whether including $c_{21}^{\rm 4PN}$ might
be useful as well as the possible extension of the TaylorF2 
approximant up to (parametrized) 5.5PN accuracy, using the 
parametrized 5PN and 5.5PN terms we derived in Eqs.~\eqref{eq:hF5PN}-\eqref{eq:hF55PN}.

\subsection{Preliminary discussion of the spinning case}
\label{sec:spinning}
Finally,  we have also driven a preliminary exploration of the spinning
case. First of all, one can incorporate the new term of Eq.~\eqref{eq:ssTail}
in the construction of the spin-aligned TaylorF2 approximant. This would add
to the known spin-orbit (non-tail) knowledge up to NNLO, the NLO spin-spin
and the LO spin cube~\cite{Bohe:2013cla,Bohe:2015ana,Marsat:2014xea,Marsat:2013wwa}.
In addition, as we did for the nonspinning case, one can similarly parametrize
higher PN spin-dependent terms. In this respect, we anticipate that, once the
computation of the factorized $\rho_{\ell m}$'s will be completed at NNLO
spin-orbit~\cite{marsat:2017} accuracy, one will have immediate access to some
of the higher order tail terms in the flux, so that it will be possible to incorporate
a lot of higher PN information in any PN-expanded approximant, though with
some dependence on parameters for the uncalculated terms in the $\rho_\lm$'s.
However, it will be necessary to investigate carefully to which extent
these terms have to be incorporated in a TaylorF2 approximant aiming
at being better than the 3.5PN one for small-mass binaries and/or to
improve current phenomenological models.
By doing a preliminary investigation of the $\hat{\Psi}'(f)$ quantity we discussed
above, it is easy to find that, when the spins are present (whatever value),
the 3.5PN one {\it is monothonically growing} after frequencies $fM\approx 0.02$,
while the ``true'' behavior captured by the IMRPhenomD model is actually qualitatively
the same as the nonspinning case discussed above, i.e. monothonically decreasing for $fM>0$.
Such growing behavior is mainly related to the $\log(\pi M f)$ term in the $\varphi_5$
coefficient of the TaylorF2 approximant, see Eq.~(B11) of Ref.~\cite{Khan:2015jqa}.
One finds that, if {\it only} the LO spin-orbit and spin-spin terms, that is Eqs.~(B9) and (B10)
of Ref.~\cite{Khan:2015jqa}, are retained in the TaylorF2, together with the same 3.5PN-accurate orbital pieces,
the behavior of $\hat{\Psi}'(f)$ is qualitatively correct, i.e. it is monotonically decresing with $fM$.
We checked this to be the case explicitly for a sample of equal-mass, equal-spin binaries
with $-1\leq \chi_i\leq 1$. As an instructive exercise, we then built a new, parametrized,
TaylorF2 approximant that is formally 4.5PN accurate in the orbital part,
but {\it only retains} LO spin-orbit and spin-spin terms. Focusing only on the equal-mass,
equal-spin case for simplicity, we find it possible, by just flexing the $c_{22}^{\rm 4PN}$
parameter, to have the 4.5PN curve consistent with the IMRPhenomD one up to higher frequencies
than those possible for the usual TaylorF2 at 3.5PN. This results illustrates at the same time
the robustness and the flexibility of our approach, although more precise and quantitative
assessment are needed and are postponed to future studies.

\section{Conclusions}
\label{sec:conclusions}
By Taylor expanding well-known analytical information
incorporated in the effective-one-body description of
BBHs, we have: (i) shown that the 4.5PN accurate tail-induced flux term
obtained by Ref.~\cite{Marchand:2016vox} using the multipolar post-minkowskian
formalism is already incorporated in the currently used EOB flux and does not need
any additional information to be obtained; (ii) we similarly obtained the
next-to-leading-order tail-induced spin-orbit term in the flux~\cite{Marsat:2013caa};
(iii) we have computed for the first time (to our knowledge) the leading-order
spin-square tail-induced term, finding perfect agreement with the test-particle
limit results of~\cite{Tagoshi:1996gh}; (iv) we have derived new
4.5PN-accurate (and higher) TaylorF2 approximants that explicitly
depend on yet uncalculated PN information. We have illustrated that,
thanks to the possibility of flexing a single parameter and focusing
on the nonspinning case for simplicity, one is able to reproduce the
derivative of the Fourier phase as provided by the IMRPhenomD model
up to $fM\approx 0.01$, while the 3.5PN approximants usually
stops at $fM\approx 0.005$.

We note in passing that our 4.5PN approximant, once complemented
by the 7.5PN accurate tidal phase of Ref.~\cite{Damour:2012yf},
should yield a better representation of the phasing for nonspinning
binary neutron stars (BNS) during the late inspiral, and up to merger,
than the 3.5PN one. We recall in fact (see e.g. Fig.~1 of
Ref.~\cite{Bernuzzi:2014owa}) that the BNS merger frequency,
estimated using NR simulations over a large sample of EOS,
approximately occurs in the range
$0.015\approx M f_{\rm mrg}^{\rm BNS}\approx 0.032$. In this
frequency range, the tuned 4.5PN TaylorF2 is closer
to the IMRPhenomD phasing than the standard 3.5PN one
(see top-left panel of Fig.~\ref{fig:hatDpsi}
and left panel of Fig.~\ref{fig:RelDiff}) and thus it might be
helpful to reduce possible systematics yielded by this latter.

In conclusion, our exploratory investigation suggests that EOB-derived,
high-order, parametrized, PN-approximants may help to produce improved
descriptions of the inspiral phasing that might be useful either for
improving current phenomenological models in the low-frequency regime,
or even as alternative detection tools for low-mass binaries.

\acknowledgments
We are grateful to T.~Damour and N.~Fornengo for discussions, as well as to L.~Blanchet, G.~Faye,
M.~Hannam, S.~Husa, T.~Marchand and P.~Schmidt for comments on the manuscript.
We also acknowledge L.~Villain for his Matlab implementation of IMRPhenomD, to S.~Husa 
for crucial help in identifying and fixing some related issues and to G.~Pratten and M.~Colleoni
for spotting an inconsistency in the published version of the manuscript that is now here corrected.

\appendix
\section{Orbital angular momentum at 4PN}
Let us briefly recall how the 4PN-accurate expression of the
orbital angular momentum given in Eq.~\eqref{eq:jorb} can be
computed by just PN expanding one of the EOB equations of motion.
Our starting point is the 4PN-accurate EOB interaction potential 
\begin{align}
 A&=1-2u+2\nu u^3+a_{4}\nu u^4\\
 &+\nu \left(\frac{64}{5}\log u+a_{5}^{c_{0}}+\nu a_{5}^{c_{1}}\right)u^5\nonumber
\end{align}
with $a_{4}=94/3-41/32\,\pi^2$,
$a_{5}^{c_{0}}=-4237/60+128/5 \gamma_{\rm E} +2275/512\,\pi^2+256/5\log(2)$
and $a_{5}^{c_{1}}=41/32\pi^2-221/6$.
The angular momentum along circular orbits is given by~\cite{Buonanno:1998gg}
\begin{align}
  \label{eq:j0}
 j_{\rm adiab}^2(u)=\frac{-A'(u)}{(u^2A(u))'},
\end{align}
where the prime stands for derivative with respect to $u$.
From the circularized EOB Hamiltonian
\begin{align}
  \label{eq:H}
 \hat{H}=\frac{1}{\nu}\sqrt{1+2\nu \left[\sqrt{ A(u)(1+j^2u^2)}-1\right]},
\end{align}
the dimensionless orbital frequency $\hat{\Omega}\equiv G M\Omega/c^3$ reads
\begin{align}
 \hat \Omega=\frac{\partial {\hat H}}{\partial{ j}},
\end{align}
which finally yields an expression of $\Omega$ as a function of $u$.
We must then invert the series to obtain, from Eq.~\eqref{eq:j0},
the orbital angular momentum in terms of the gauge-invariant
parameter $x\equiv\hat \Omega^{2/3}$ as given in
Eq.~\eqref{eq:jorb} in the main text.

\bibliography{refs20180605}

\end{document}